\documentclass[
aps,
prl,
reprint,
nofootinbib,
showkeys,
]{revtex4-1}

\usepackage[utf8]{inputenc}
\usepackage{amsmath}
\usepackage{amsfonts}
\usepackage{amssymb}
\usepackage{graphicx}
\usepackage{graphics}
\usepackage{lipsum}
\usepackage{dcolumn}
\usepackage[dvipsnames]{xcolor,colortbl}
\definecolor{Gray}{gray}{0.85}
\usepackage{txfonts}
\usepackage[T1]{fontenc}
\usepackage[hidelinks]{hyperref}
\hypersetup{
	colorlinks   = true, 
	urlcolor     = blue, 
	linkcolor    = blue, 
	citecolor    = blue 
}

\begin{document}
	\renewcommand{\figureautorefname}{Fig.}
	\renewcommand{\equationautorefname}{Eq.}
	
	\newcommand{\defColorOne}{red}
	\newcommand{\dcM}{\color{\defColorOne}}

	\newcommand{\defColorTwo}{Red}
	\newcommand{\dcL}{\color{\defColorTwo}}
	
	\newcommand{\defColorThree}{blue}
	\newcommand{\jr}{\color{\defColorThree}}

	\title{Ultra-flat twisted superlattices in 2D heterostructures}
	\author{Márton \surname{Szendrő}} 
	\author{Péter \surname{Süle}}
	\author{Gergely \surname{Dobrik}} 
	\author{Levente \surname{Tapasztó}}
	\affiliation{Institute of Technical Physics and Materials Science (MFA), Centre for Energy Research, Hungarian Academy of Sciences, 1525 Budapest, P.O. Box 49, Hungary}
	
	\begin{abstract}
		Moiré-superlattices are ubiquitous in 2D heterostructures, strongly influencing their electronic properties. They give rise to new Dirac cones and are also at the origin of the superconductivity observed in magic-angle bilayer graphene. The modulation amplitude (corrugation) is an important yet largely unexplored parameter
		in defining the properties of 2D superlattices. The generally accepted view is that the corrugation
		monotonically decreases with increasing twist angle, while its effects on the electronic structure
		diminish as the layers become progressively decoupled. Here we found by lattice relaxation of around 8000 different Moiré-superstructures using high scale Classical Molecular Simulations combined with analytical calculations, that even a small amount of strain can substantially change this picture, giving rise to
		more complex behavior of superlattice corrugation as a function of twist angle. One of the most surprising findings is the emergence of an ultra-flat phase that can be present for arbitrary small twist angle
		having a much lower corrugation level than the decoupled phase at large angles. A possible experimental realization of the ultra-flat state is revealed by Scanning Tunneling Microscopy (STM) investigations of the graphene/graphite system.  
	\end{abstract}
	
	\date{\today}
	\maketitle
	
	\section{Introduction}
	
	When 2D materials are layered on top of each other an interference pattern called the Moiré-pattern is formed due to the lattice mismatch and relative rotation between the layers. The Morié-pattern acts as an additional external periodic potential, which perturbs the electronic states of the layers leading to a large variety of physical phenomena such as secondary Dirac cones \cite{ponomarenko2013cloning,yankowitz2012emergence,dean2013hofstadter,hunt2013massive}, Hofstadter's butterfly \cite{dean2013hofstadter,hunt2013massive,yu2014hierarchy,wang2015evidence}, Brown-Zak oscillations \cite{kumar2017high}, valley polarized currents \cite{gorbachev2014detecting}, Moiré-excitons \cite{rivera2015observation,rivera2016valley,wu2018theory,yu2017moire,tran2019evidence}, and recently superconductive \cite{cao2018unconventional} and Mott-insulating states \cite{cao2018correlated}. The corrugation stemming from the Moiré-pattern is often neglected in theoretical calculations as the periodic potential itself is sufficient to account for the main aspects of these phenomena. However, the way 2D heterostructures relax strain through out-of-plane deformation can highly influence the properties of such systems, enabling us to gain more detailed insights into the origins and rich physics of these phenomena. Experimental evidence for the out-of-plane deformation of a freestanding graphene/h-BN single layer 2D heterostructure has been provided by transmission electron microscopy investigations \cite{argentero2017unraveling}.
	 
	Corrugated Moiré-patterns have nonuniform strains which lead to pseudo-magnetic fields and consequently an energy gap in the graphene in the range of 20-60 meV \cite{neek2014graphene,nam2017lattice}. Heavily corrugated Moiré in gr/Ru(0001) shows localized states close to the Fermi-level \cite{stradi2012electron}. It has been shown that lattice relaxation, (which leads to a corrugated surface) is an important parameter to understand the origin of the gap in the gr/h-BN system \cite{jung2015origin,yankowitz2018dynamic}. The importance of the relaxation was also pointed out for low angle twisted bilayer graphene (TBLG), where an energy gap of up to 20 meV opens at the superlattice subband edge \cite{nam2017lattice,choi2018strong}, which is also found in experiments \cite{cao2018correlated,cao2016superlattice} and is needed to fully describe the Mott-physics of the system. Small-angle TBLG under interlayer bias also shows topologically protected helical modes in stacking boundaries \cite{san2013helical,tong2017topological,huang2018topologically}. Increasing the corrugation of the stacking boundaries cause quantum-well-like gapped states to abound, which give rise to robust conductance channels along the corrugated boundaries \cite{pelc2015topologically}. There is also a quest to synthesize ultra-flat graphene which is driven by the fact, that height fluctuations can cause charge inhomogeneities in the graphene which degrade its transport properties \cite{lui2009ultraflat,xue2011scanning}. On crystalline atomically flat surfaces the main height fluctuation arises from the height modulation of the Moiré-pattern.
	
	Despite the role of the Moiré-corrugation in a wide range of phenomena it is still a largely unexplored field and is lacking a comprehensive theoretical description. The commonly accepted picture is that the corrugation is a monotonically decreasing function of the twist angle. This is generally due to the decreasing Moiré-wavelength with increasing twist angle. Moiré-hills with smaller wavelength would have a higher aspect ratio, and therefore a higher radius of curvature, which would heavily increase the elastic energy if the corrugation were not reduced to compensate for this effect. Few theoretical calculations have been developed to describe the amplitude modulation \cite{neek2014membrane,gao2011effect,aitken2010effects,thuermer2015real}, however, the substrate was considered to be rigid and/or perfectly flat in these works. These assumptions hinder some of the complex behavior of the Moiré-corrugation as we will demonstrate in this work. 
	
	\begin{figure}
	\begin{center}
		\includegraphics[width=0.515\textwidth]{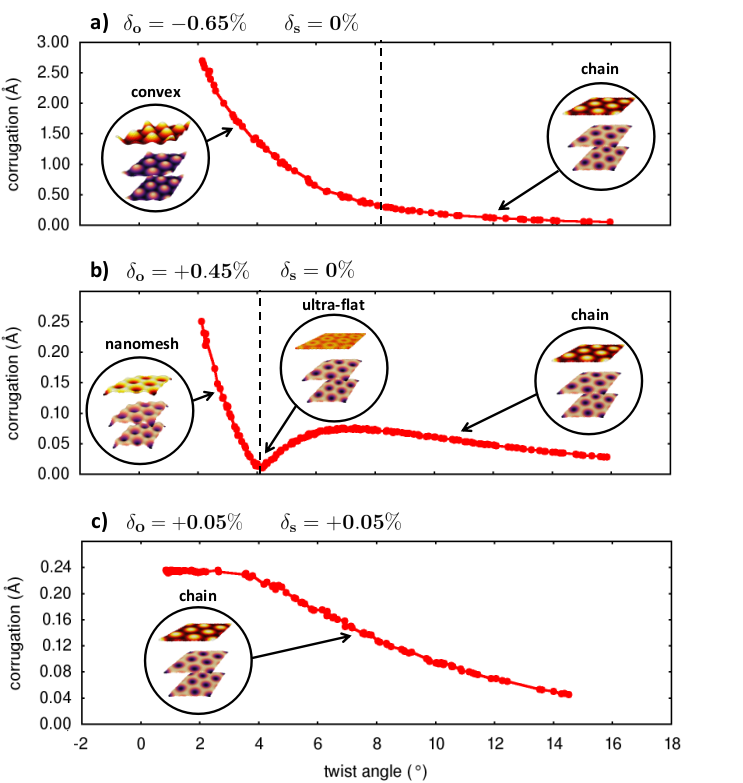}
		\caption{\label{fig:1} Moiré-superlattice corrugations of the top rotated graphene layer for three differently strained T5LG (one twisted graphene layer above four Bernal stacked graphene layers) heterostructures (red dots) from molecular mechanics simulations. Each point represents a commensurate Moiré-superlattice. a) The overlayer is externally compressed with $\sim$0.65\% during minimization which leads to a convex phase for angles below 8.2$^{\circ}$ and a chain phase for angles above. The phase transition, indicated by dashed lines, is smooth, and the whole curve is monotonically decreasing. b) The overlayer is stretched with $\sim$0.45\% strain. This cause a nanomesh phase to emerge for angles below 4$^{\circ}$, and a chain phase for angles above. Near the phase transition (dashed line), the corrugation vanishes and an ultra-flat phase appears.  c) Both the overlayer and the substrate are stretched with a small 0.05\% strain. No phase-transition occurs, each point corresponds to a chain structure. In the range of 0-4$^{\circ}$ the corrugation is closely constant (plateau-effect). The topographies of the top three layers of the corresponding Moiré-phases are displayed in bubbles.}
	\end{center}
	\end{figure}

	We used high scale Classical Molecular Mechanics simulations for relaxating approximately 8000 differently strained and twisted Moiré-supercells. Our model systems were the twisted trilayer graphene (T3LG) and twisted five-layer graphene (T5LG), both consisting of Bernal stacked graphene layers with one rotated layer on top. 
	Such few-layer graphene structures turned out to be ideal model systems for studying topographical phases of Moiré-patterns. We show that by including strain in the system three typical Moiré-topographies (Moiré-phases) can be present in 2D heterostructures.  The existence of various Moiré-phases enriches the corrugation physics causing that even small strain perturbations lead to complex corrugation behavior. 
	
	\section{Results and Discussion}
	
	To asses the effect of external strain on the Moiré-corrugation we performed lattice relaxations with differently strained and twisted T3LG and T5LG heterostructures. 
	We used the T3LG structure  for the more computation-intensive calculations. 
	Twisted few-layer graphene structures proved to be suitable model systems due to the following reasons: (i) we could use DFT validated force fields for these systems (see {\jr Fig. S2}), (ii) it also captures the asymmetry characteristic of 2D layer/substrate systems as opposed to bilayer graphene, (iii) every Moiré-phase is reachable for reasonable strains in a wide range of twist angles (as we will show later). 
	
	We applied homogeneous biaxial strain $\delta$ independently to the overlayer ($\delta_{o}$) i.e. the top layer and to the substrate ($\delta_{s}$) i.e. the remaining layers. Firstly, we calculated the Moiré-corrugation of relaxed T5LG superstructures as a function of twist angle in the range of $\alpha = 2^{\circ}-16^{\circ}$  for three distinct strain configurations (see \autoref{fig:1}).   
	The most striking finding is that the strain crucially influences the behavior of the corrugation curves. In the first case (\autoref{fig:1} a) the overlayer was compressed by $\delta_{o} = -0.65 \%$ but the substrate was left strain-free $\delta_{s} = 0 \%$. The corrugation curve of the overlayer shows the expected monotonically decreasing behavior. In the second case (\autoref{fig:1} b) the overlayer was stretched by $\delta_{o} = +0.45 \%$, still leaving the substrate with no external strain. The overlayer corrugation is no longer monotonically decreasing in the whole $\alpha$ range, rather it tends to vanish at around $4^{\circ}$, then increasing between $4^{\circ}-7^{\circ}$ before reaching again the decreasing tendency characteristic to large twist angles. As for the third case (\autoref{fig:1} c) both the substrate and the overlayer were stretched by a small amount of $\delta_{o} = +0.05\%$, $\delta_{s} = +0.05\%$, which leads to a curve having a plateau in the interval of $0^{\circ}-4^{\circ}$ and then monotonically decreasing in the rest.
	
	The significant differences of the corrugation curves triggered by very mild strain levels hint an intriguing underlying mechanism to be revealed. Also for a particular $\alpha = 4^{\circ}$, $\delta_{o} = +0.45\%$, $\delta_{s} = 0 \%$ parameter set an ultra-flat state surprisingly appears in the overlayer. In order to get a better insight, we investigated in details the resulting topographies. 
	We found that during rotation, not only the periodicity and corrugation but also the morphology of the Moiré-superlattice can change. Such Moiré-phase transitions are marked by vertical dashed lines in \autoref{fig:1}. Sometimes these phase transitions do not manifest themselves in corrugation change (\autoref{fig:1} a), while in other cases the corrugation can drastically alter around them (\autoref{fig:1} b). The corresponding topographies of the top three layers of the phases are depicted in bubbles in \autoref{fig:1}.
	
	In 2D heterostructures, corrugation can develop in both upper and lower layers. The Moiré-amplitude in the substrate can be in phase with the overlayer amplitude (bending modes), or in anti-phase (breathing mode). There are two bending modes (convex, nanomesh) and one breathing (chain), which we call together \emph{Moiré-phases}. In the \emph{convex-phase} the AA regions of the Moiré-pattern are forming protrusions, while the remaining two AB/BA stackings are depressions (\autoref{fig:1} a). The \emph{nanomesh-phase} is the inverse of this: the AA regions are depressions and AB/BA stackings are the protrusions (\autoref{fig:1} b). In the convex and nanomesh phases, both the overlayer and the substrate has the same type of topography (bending-modes). However, the third phase, the chain-phase (\autoref{fig:1} a,b,c) is qualitatively different as the overlayer and the substrate are closely mirror images of each other having a convex topography in the overlayer and a concave in the substrate (breathing mode). 
	
\begin{figure*}
		\begin{center}
			\includegraphics[width=0.9\textwidth]{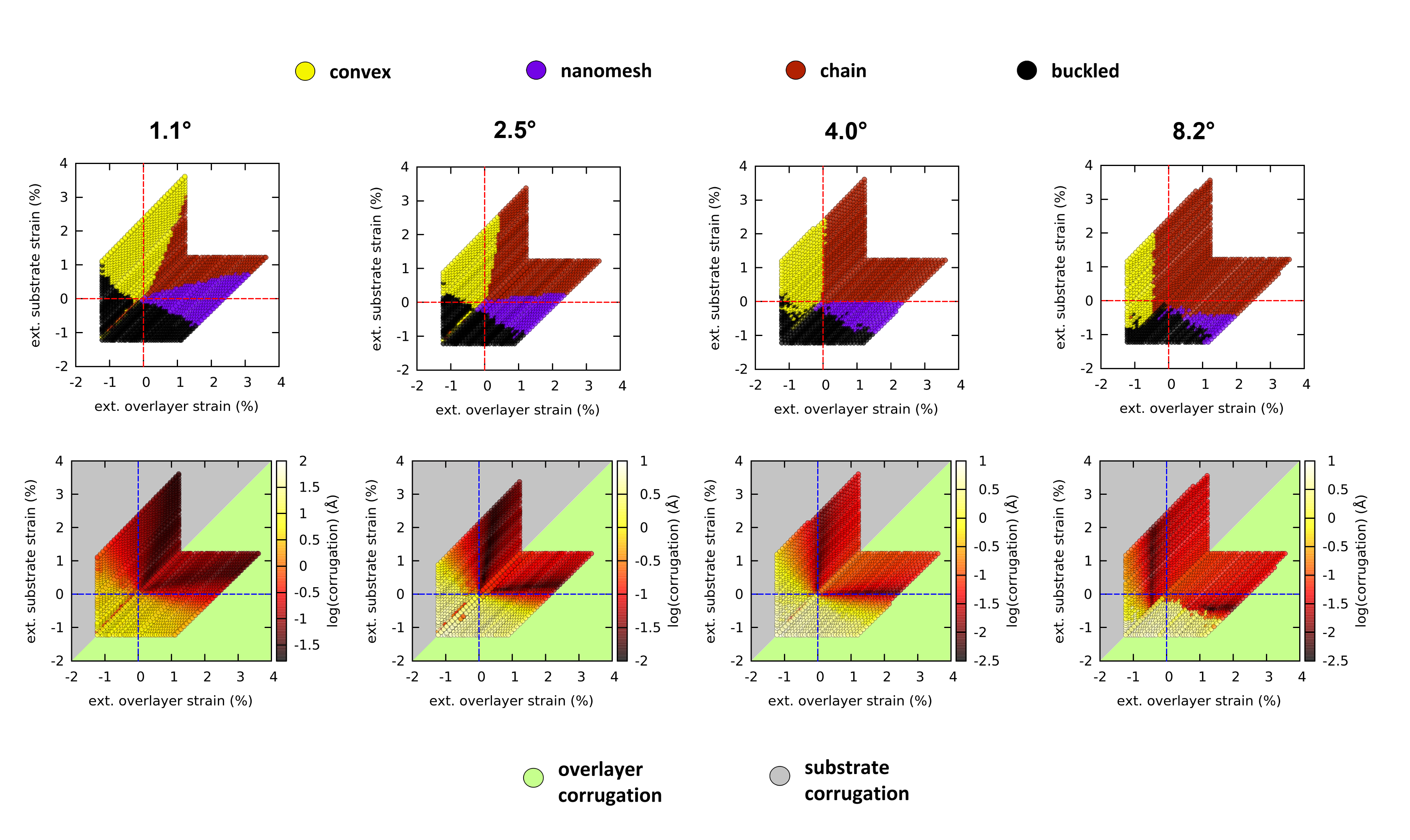}
			\caption{\label{fig:2} \textit{Top row:} Moiré-phases of the twisted trilayer graphene (T3LG) system for different twist angles in the space of externally applied homogeneous strains of the overlayer and of the substrate (phase-maps). A single point in a map represents a relaxed T3LG commensurate Moiré-superlattice. The corresponding Moiré-phases are indicated with different colors (yellow - convex, purple - nanomesh, red - chain). As the twist angle increases the phase boundaries move, and the chain phase becomes dominant. \textit{Bottom row:} The same phase-maps but colored by the corrugation of the relaxed Moiré-superlattices. The dark areas show the ultra-flat states around the phase boundaries. On the side of the green shaded area, the overlayer corrugation is shown, while on the gray side the substrate corrugation is displayed, in order to show the ultra-flat states both in the overlayer and in the substrate. }
		\end{center}
	\end{figure*}

	The results above clearly reveal that for different $\alpha$, $\delta_{o}$, $\delta_{s}$ parameters different Moiré-phase realizations occur, and even ultra-flat states can emerge near some of the phase transitions. To capture all the Moiré-phases emerging in our system, as well as the conditions necessary for the realization of ultra-flat states, the whole ($\alpha$, $\delta_{o}$, $\delta_{s}$) phase space has to be studied by examining the resulting topographies and corrugations for each point in this space. Therefore we created phase-maps of the Moiré-phases in the space of $(\delta_{o},\delta_{s})$ for four different twist angles $\alpha=$1.1$^{\circ}$, 2.5$^{\circ}$, 4.0$^{\circ}$, 8.2$^{\circ}$. The sampling of the whole phase-space needs a large number of superstructure relaxations, therefore we used the fewer layer T3LG structure for this task instead of the T5LG. For each phase-map with a definite $\alpha$, we searched for about 60 Moiré T3LG supercells having different $\delta_{o},\delta_{s}$ combinations. Then we stretched each supercell structure in steps of $\delta \approx 0.085 \% $. This way we ended up with approximately 1800-2000 Moiré-supercells for each phase-map. Every Moiré T3LG superstructure was relaxed with a fixed simulation cell using in-plane periodic boundary conditions and the same setup as in the case of \autoref{fig:1}. The type of morphology and corrugation was analyzed after the relaxations. The results of the $\sim$8000 Moiré-superstructures can be seen in \autoref{fig:2}.  
	
	Three Moiré-phases are present in the phase-maps for the examined angles (\autoref{fig:2} top row). Also for compressive strains higher than a threshold value, the Moiré-pattern is absent, and the whole structure is buckled due to the buckling instability of graphene while having a much larger corrugation. We denoted the buckled states with black. Incidentally, the buckled states are twist-angle dependent: increased buckling threshold can be seen in the convex phase for higher twist angles. There are three phase-boundaries between the Moiré-phases. Two out of the three are more extended, namely the nanomesh-chain boundary and the convex-chain boundary. The convex-nanomesh boundary is small and is vanishing for larger twist angles as this boundary is completely merging with the buckled states. Furthermore, there is a triple point where the three phases coexist. The fascinating thing is that the position of the strain-free $\delta_{o} = 0$, $\delta_{s} = 0$ state is very close to the triple point for angles between $0^{\circ}-4^{\circ}$. What this means is that  very delicate strain fluctuations can lead to very different phase realizations, therefore the $\delta_{o} = 0$, $\delta_{s} = 0$ state is very sensitive from this point of view. Not only it is sensitive for the realization of various Moiré-phases but from a corrugation perspective. In the bottom row of \autoref{fig:2}, we show the corrugation value of each relaxed Moiré-supercell. When stretching the whole system gradually the corrugation is decreasing as it is expected; however, for a certain set of finite $\delta_{o},\delta_{s}$, the corrugation drops rapidly close to zero (dark regions in the bottom row of \autoref{fig:2}). These are the already discussed ultra-flat states. They run along the convex-chain and the nanomesh-chain phase boundaries. The main difference in the two boundaries is that in the former case the substrate corrugation is zero, while in the latter the overlayer. This is the reason why we could not see ultra-flat states in \autoref{fig:1} a) as these states occur in the substrate. The emergence of ultra-flat states can be linked to the convexity change of the topographies. The convexity of the overlayer is positive both in
	the convex and in the chain-phase, therefore the transition between these two phases can be continuous in the corrugation. However, when moving from nanomesh to chain the convexity changes sign progressively in the overlayer implying a particular twist angle in between, where the corrugation has to be zero. Nonetheless, this effect takes place in the substrate inversely: the change of the convexity occurs during the convex-chain transition with consequent ultra-flat states appearing in the substrate, but not in the overlayer (see also {\jr Fig. S7}).
	
	The ultra-flat states go along the boundaries, however, they do not meet in the middle at $\delta_{o} = 0$, $\delta_{s} = 0$. Now it can be clearly seen, that around the $\delta_{o} = 0$, $\delta_{s} = 0$ point, the corrugation strongly fluctuates. It means that in realistic samples even small strain fluctuations can be crucial in determining the topography of the system, which can take shape from being ultra-flat to highly corrugated, or with convexity from positive to negative. This is a very important finding for the practical design of 2D heterostructures, pointing out the crucial role of even mild strains that have to be controlled  very accurately for reliably engineering the morphology of  Moiré-superlattices in 2D heterostructures.
	
	The phase boundaries (convex-chain, nanomesh-chain) are also moving as the twist angle changes. With increasing twist angle we see the chain phase being more dominant on the phase map. The movement of the boundaries takes place in two distinct stages. For smaller twist angles ($0^{\circ}-4^{\circ}$) only the angle between the phase boundaries are changing and the boundaries are rotating. For larger angles ($>4^{\circ}$) the phase boundaries are perpendicular to each other, and their movement consists of pure translation. The movement of the phase boundaries makes it possible for a system to go under a phase transition while altering the twist angle as it was shown in \autoref{fig:1}.

		The interpretations of the results above can be attained through energetic considerations. The Moiré-morphology is defined by the balance between the elastic energy and the vdW adhesion. The elastic energy stemming from the corrugation $\xi$ with a Moiré-wavelength $\lambda$ is a power series of $\xi/\lambda$ {\cite{aitken2010effects}}. The vdW contribution is determined by the interlayer separation of the AA stacking $d_{1}$ and the AB/BA stackings $d_{2}$. The corrugation in the overlayer $\xi_{o}$ and substrate $\xi_{s}$ are coupled through vdW interaction. In particular, it is easy to show, that $\xi_{o} = \xi_{s} + \Delta d$ (see {\jr Fig. S3}), where $\Delta d = d_{1} - d_{2}$, which is a strict constraint on the geometry of the phases. It can be seen (see Section I. of Suppl. Mat.), that in the convex phase $\xi_{o} \geq \Delta d$ and $\xi_{o} > \xi_{s}$, while $\xi_{s} \geq \Delta d$ and $\xi_{o} < \xi_{s}$ in the nanomesh phase. In the chain phase $0 \leq \xi_{o} \leq \Delta d$, $0 \leq \xi_{s} \leq \Delta d$. These geometric properties have a strong impact on the realization of the phases. For example, a more corrugated overlayer than the substrate can be realized only in the convex or the chain phase.  In the case when the overlayer does not want to corrugate as much as the substrate then the nanomesh or the chain phase is favorable. When the corrugation is equally unfavorable for the overlayer and for the substrate, we expect the chain phase to emerge. Also as the elastic energy goes with the powers of $\xi/\lambda$, the energy increment with a unit $d\xi$ will be much bigger for larger twist angles. Therefore, there is a tendency to divide the total corrugation evenly between the substrate and the overlayer as the twist angle increases. This geometry, however, can only be realized in the chain phase. Consequently, with increasing twist angle we expect to see the chain phase being more dominant on the phase maps. Moreover, for small twist angles, $\xi/\lambda$ tends to zero and the vdW adhesion becomes prominent, while for large twist angles the elastic energy is the major energy term (see {\jr Fig. S9}). The plateau on \autoref{fig:1} c) is the consequence of the special geometry of the chain-phase combined with these two energetic regimes ({\jr Fig. S9}). The boundary movements (rotation, translation) also coincides with these two regimes. The detailed analysis of the results on this energetic basis can be found in the Supplementary Material, where we established the general picture of the Moiré-phases, and also developed an analytic one-dimensional string model. Our analytic results ({\jr Fig. S7-S12}) are in good qualitative agreement with the outcome of the molecular simulations.

\begin{figure*}
	\begin{center}
		\includegraphics[width=0.8\textwidth]{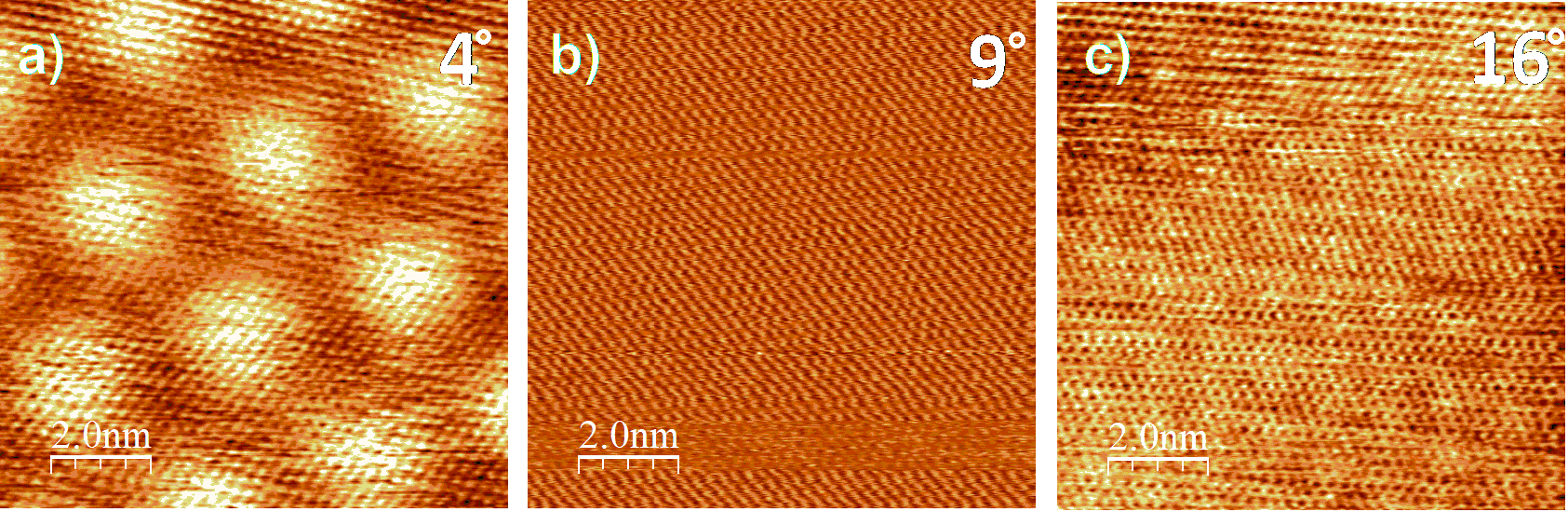}
		\caption{\label{fig:3}Topographic scanning tunneling microscope images of graphene layers deposited on top of a graphite substrate for various relative rotation angles. The STM image in panel (b) reveals an ultra-flat state, much smoother than observed even for high rotation angles (c). The experimental conditions
		for image acquisition (I tunel = 1nA, U bias = 200mV), data processing, and graphic display parameters are the same for all panels.}
	\end{center}
\end{figure*}

The experimental validation of our theoretical predictions is rather difficult as we cannot fully control the amount of heterostrain emerging in the system. Nevertheless, we have prepared and analyzed graphene layers deposited at various rotation angles on graphite substrates. We were able to measure the rotation angle, Moiré-periodicity, as well as the Moiré-amplitude for various rotation angles. Although due to the lack of control over the heterostrain, we cannot reliably plot curves similar to the theoretical plots in {\autoref{fig:1}}, the general tendency observed was  that the corrugation is high at small rotation angles and low at high angles, which is in accordance with the previous expectations. However, we were also able to find a graphene flake where the corrugation was very low almost undetectable at an intermediate  (9$^{\circ}$) rotation angle. We made sure that this is not an artifact, as the atomic corrugation was as expected (no resolution issue). Furthermore, a very faint superlattice signature could still be observed in larger area images, ensuring that the graphene overlayer is not decoupled by contamination from the graphite substrate. The experimentally determined average Moiré-amplitudes for the three rotation angles shown in \autoref{fig:3} are as follows: 
	0.32 \AA $\pm$ 0.02 \AA $~$(4.2$^{\circ}$ $\pm$ 0.1$^{\circ}$); 
	0.05 \AA $\pm$ 0.02\AA $~$(9.7$^{\circ}$ $\pm$ 0.1$^{\circ}$); 
	0.12 \AA $\pm$ 0.02 \AA $~$(16$^{\circ}$ $\pm$ 0.5$^{\circ}$).  
Consequently, the graphene flake displayed in \autoref{fig:3} b) can be regarded as an experimental realization of an ultra-flat state predicted theoretically, although the exact parameters (heterostrain values) could not be directly inferred. 
Therefore the direct quantitative comparison of the experimentally found ultra-flat state with the simulations is not straightforward as the unknown strain values essentially influence the twist angle at which the ultra-flat states emerge as it was shown in \autoref{fig:2}. The exact shape of the vdW interactions also affects the twist angle value of the ultra-flat state (see {\jr Fig. S8 d}). These effects can be at the origin of the quantitative discrepancy between the theoretically predicted and experimentally observed twist angles for the ultra-flat states.
Changes in the Moiré-morphologies (i.e. from convex to concave) have also been observed experimentally ({\jr Fig. S1}) on the same graphene flake (for the same rotation angle), which can be induced by the locally varying strain, in agreement with the predicted high strain-sensitivity of the Moiré-phase realizations. The high strain-sensitivity 
of Moiré-superlattices is also supported by the experimental observation 
of the coexistence of the convex and nanomesh morphologies of graphene 
superlattices on Au(111) \cite{sule2015nanomesh}. 
	
In conclusion, we showed using Molecular Mechanics simulations that strain and twist angle alteration can induce phase transitions in twisted 2D superlattices between three typical topographical Moiré-phases. On phase boundaries, ultra-flat states were identified where the corrugation is almost zero either in the overlayer or in the substrate, but not both. These ultra-flat states were shown to be also present for small twist angles, where conventionally a large corrugation is expected. We also showed that for smaller twist angles ($0^{\circ}-4^{\circ}$) the strain-free equilibrium point is critical and even very small strains can induce significant changes in the corrugation and in the  morphology of the superlattices. The results and concepts developed here are expected to be essential for the rational design of twisted 2D superlattices and highlight the critical role played by externals strains in defining the nanoscale morphology of such 2D heterostructures.

\section{Methods}

The classical molecular energy minimizations were carried out using the LAMMPS code \cite{plimpton1995fast}. In-plane periodic boundary conditions were used. During the relaxation the size of the simulation cell was fixed, which ensured through the periodic boundary conditions that the pre-applied strain can not be fully eliminated, and that $\delta_{o}, \delta_{s}$  is an additional constraint on the energy minimum. The long range bond order potential was utilized \cite{los2003intrinsic} for carbon-carbon interactions within a single carbon layer. The weak Van der Walls forces between the carbon layers were modeled using the Kolmogorov-Crespi potential \cite{kolmogorov2005registry}. We performed geometry relaxations on commensurate Moiré-supercells using Hessian-free Truncated Newton algorithm implemented in LAMMPS as the \emph{hftn} algorithm. Each supercell used in this paper was found by our script iterating through highly ordered commensurate cells as described in \cite{merino2011strain,meng2012multi} to match the criterion of the cell having a definite $\alpha$, $\delta_{o}$, $\delta_{s}$ property. Due to the fact that for a given arbitrary $\alpha$, $\delta_{o}$, $\delta_{s}$ there not necessarily exists a corresponding supercell, the $\alpha$, $\delta_{o}$, $\delta_{s}$ values may display minor fluctuations that limit the angle and strain resolution of our data; nevertheless do not affect their interpretation.   

\section{Data Availability}
The data supporting the present work are available from the
corresponding author upon reasonable request.

\section{Acknowledgments}
We acknowledge support form NanoFab2D ERC starting grant and NIIF for awarding us access to resource based in Hungary at Budapest and Debrecen.

\section{Author Contributions}
M.SZ. performed the computational and theoretical calculations. G.D. carried out the STM investigations. P.S. and L.T. supervised the project. M.SZ. and L.T. wrote the paper. All authors discussed the results and commented on the manuscript.  

\section{Additional Information}
\textbf{Competing interests:} The authors declare no competing interests.   


\begin{thebibliography}{39}%
\makeatletter
\providecommand \@ifxundefined [1]{%
 \@ifx{#1\undefined}
}%
\providecommand \@ifnum [1]{%
 \ifnum #1\expandafter \@firstoftwo
 \else \expandafter \@secondoftwo
 \fi
}%
\providecommand \@ifx [1]{%
 \ifx #1\expandafter \@firstoftwo
 \else \expandafter \@secondoftwo
 \fi
}%
\providecommand \natexlab [1]{#1}%
\providecommand \enquote  [1]{``#1''}%
\providecommand \bibnamefont  [1]{#1}%
\providecommand \bibfnamefont [1]{#1}%
\providecommand \citenamefont [1]{#1}%
\providecommand \href@noop [0]{\@secondoftwo}%
\providecommand \href [0]{\begingroup \@sanitize@url \@href}%
\providecommand \@href[1]{\@@startlink{#1}\@@href}%
\providecommand \@@href[1]{\endgroup#1\@@endlink}%
\providecommand \@sanitize@url [0]{\catcode `\\12\catcode `\$12\catcode
  `\&12\catcode `\#12\catcode `\^12\catcode `\_12\catcode `\%12\relax}%
\providecommand \@@startlink[1]{}%
\providecommand \@@endlink[0]{}%
\providecommand \url  [0]{\begingroup\@sanitize@url \@url }%
\providecommand \@url [1]{\endgroup\@href {#1}{\urlprefix }}%
\providecommand \urlprefix  [0]{URL }%
\providecommand \Eprint [0]{\href }%
\providecommand \doibase [0]{http://dx.doi.org/}%
\providecommand \selectlanguage [0]{\@gobble}%
\providecommand \bibinfo  [0]{\@secondoftwo}%
\providecommand \bibfield  [0]{\@secondoftwo}%
\providecommand \translation [1]{[#1]}%
\providecommand \BibitemOpen [0]{}%
\providecommand \bibitemStop [0]{}%
\providecommand \bibitemNoStop [0]{.\EOS\space}%
\providecommand \EOS [0]{\spacefactor3000\relax}%
\providecommand \BibitemShut  [1]{\csname bibitem#1\endcsname}%
\let\auto@bib@innerbib\@empty
\bibitem [{\citenamefont {Ponomarenko}\ \emph {et~al.}(2013)\citenamefont
  {Ponomarenko}, \citenamefont {Gorbachev}, \citenamefont {Yu}, \citenamefont
  {Elias}, \citenamefont {Jalil}, \citenamefont {Patel}, \citenamefont
  {Mishchenko}, \citenamefont {Mayorov}, \citenamefont {Woods}, \citenamefont
  {Wallbank} \emph {et~al.}}]{ponomarenko2013cloning}%
  \BibitemOpen
  \bibfield  {author} {\bibinfo {author} {\bibfnamefont {L.}~\bibnamefont
  {Ponomarenko}}, \bibinfo {author} {\bibfnamefont {R.}~\bibnamefont
  {Gorbachev}}, \bibinfo {author} {\bibfnamefont {G.}~\bibnamefont {Yu}},
  \bibinfo {author} {\bibfnamefont {D.}~\bibnamefont {Elias}}, \bibinfo
  {author} {\bibfnamefont {R.}~\bibnamefont {Jalil}}, \bibinfo {author}
  {\bibfnamefont {A.}~\bibnamefont {Patel}}, \bibinfo {author} {\bibfnamefont
  {A.}~\bibnamefont {Mishchenko}}, \bibinfo {author} {\bibfnamefont
  {A.}~\bibnamefont {Mayorov}}, \bibinfo {author} {\bibfnamefont
  {C.}~\bibnamefont {Woods}}, \bibinfo {author} {\bibfnamefont
  {J.}~\bibnamefont {Wallbank}},  \emph {et~al.},\ }\href@noop {} {\bibfield
  {journal} {\bibinfo  {journal} {Nature}\ }\textbf {\bibinfo {volume} {497}},\
  \bibinfo {pages} {594} (\bibinfo {year} {2013})}\BibitemShut {NoStop}%
\bibitem [{\citenamefont {Yankowitz}\ \emph {et~al.}(2012)\citenamefont
  {Yankowitz}, \citenamefont {Xue}, \citenamefont {Cormode}, \citenamefont
  {Sanchez-Yamagishi}, \citenamefont {Watanabe}, \citenamefont {Taniguchi},
  \citenamefont {Jarillo-Herrero}, \citenamefont {Jacquod},\ and\ \citenamefont
  {LeRoy}}]{yankowitz2012emergence}%
  \BibitemOpen
  \bibfield  {author} {\bibinfo {author} {\bibfnamefont {M.}~\bibnamefont
  {Yankowitz}}, \bibinfo {author} {\bibfnamefont {J.}~\bibnamefont {Xue}},
  \bibinfo {author} {\bibfnamefont {D.}~\bibnamefont {Cormode}}, \bibinfo
  {author} {\bibfnamefont {J.~D.}\ \bibnamefont {Sanchez-Yamagishi}}, \bibinfo
  {author} {\bibfnamefont {K.}~\bibnamefont {Watanabe}}, \bibinfo {author}
  {\bibfnamefont {T.}~\bibnamefont {Taniguchi}}, \bibinfo {author}
  {\bibfnamefont {P.}~\bibnamefont {Jarillo-Herrero}}, \bibinfo {author}
  {\bibfnamefont {P.}~\bibnamefont {Jacquod}}, \ and\ \bibinfo {author}
  {\bibfnamefont {B.~J.}\ \bibnamefont {LeRoy}},\ }\href@noop {} {\bibfield
  {journal} {\bibinfo  {journal} {Nature Physics}\ }\textbf {\bibinfo {volume}
  {8}},\ \bibinfo {pages} {382} (\bibinfo {year} {2012})}\BibitemShut {NoStop}%
\bibitem [{\citenamefont {Dean}\ \emph {et~al.}(2013)\citenamefont {Dean},
  \citenamefont {Wang}, \citenamefont {Maher}, \citenamefont {Forsythe},
  \citenamefont {Ghahari}, \citenamefont {Gao}, \citenamefont {Katoch},
  \citenamefont {Ishigami}, \citenamefont {Moon}, \citenamefont {Koshino} \emph
  {et~al.}}]{dean2013hofstadter}%
  \BibitemOpen
  \bibfield  {author} {\bibinfo {author} {\bibfnamefont {C.~R.}\ \bibnamefont
  {Dean}}, \bibinfo {author} {\bibfnamefont {L.}~\bibnamefont {Wang}}, \bibinfo
  {author} {\bibfnamefont {P.}~\bibnamefont {Maher}}, \bibinfo {author}
  {\bibfnamefont {C.}~\bibnamefont {Forsythe}}, \bibinfo {author}
  {\bibfnamefont {F.}~\bibnamefont {Ghahari}}, \bibinfo {author} {\bibfnamefont
  {Y.}~\bibnamefont {Gao}}, \bibinfo {author} {\bibfnamefont {J.}~\bibnamefont
  {Katoch}}, \bibinfo {author} {\bibfnamefont {M.}~\bibnamefont {Ishigami}},
  \bibinfo {author} {\bibfnamefont {P.}~\bibnamefont {Moon}}, \bibinfo {author}
  {\bibfnamefont {M.}~\bibnamefont {Koshino}},  \emph {et~al.},\ }\href@noop {}
  {\bibfield  {journal} {\bibinfo  {journal} {Nature}\ }\textbf {\bibinfo
  {volume} {497}},\ \bibinfo {pages} {598} (\bibinfo {year}
  {2013})}\BibitemShut {NoStop}%
\bibitem [{\citenamefont {Hunt}\ \emph {et~al.}(2013)\citenamefont {Hunt},
  \citenamefont {Sanchez-Yamagishi}, \citenamefont {Young}, \citenamefont
  {Yankowitz}, \citenamefont {LeRoy}, \citenamefont {Watanabe}, \citenamefont
  {Taniguchi}, \citenamefont {Moon}, \citenamefont {Koshino}, \citenamefont
  {Jarillo-Herrero} \emph {et~al.}}]{hunt2013massive}%
  \BibitemOpen
  \bibfield  {author} {\bibinfo {author} {\bibfnamefont {B.}~\bibnamefont
  {Hunt}}, \bibinfo {author} {\bibfnamefont {J.}~\bibnamefont
  {Sanchez-Yamagishi}}, \bibinfo {author} {\bibfnamefont {A.}~\bibnamefont
  {Young}}, \bibinfo {author} {\bibfnamefont {M.}~\bibnamefont {Yankowitz}},
  \bibinfo {author} {\bibfnamefont {B.~J.}\ \bibnamefont {LeRoy}}, \bibinfo
  {author} {\bibfnamefont {K.}~\bibnamefont {Watanabe}}, \bibinfo {author}
  {\bibfnamefont {T.}~\bibnamefont {Taniguchi}}, \bibinfo {author}
  {\bibfnamefont {P.}~\bibnamefont {Moon}}, \bibinfo {author} {\bibfnamefont
  {M.}~\bibnamefont {Koshino}}, \bibinfo {author} {\bibfnamefont
  {P.}~\bibnamefont {Jarillo-Herrero}},  \emph {et~al.},\ }\href@noop {}
  {\bibfield  {journal} {\bibinfo  {journal} {Science}\ }\textbf {\bibinfo
  {volume} {340}},\ \bibinfo {pages} {1427} (\bibinfo {year}
  {2013})}\BibitemShut {NoStop}%
\bibitem [{\citenamefont {Yu}\ \emph {et~al.}(2014)\citenamefont {Yu},
  \citenamefont {Gorbachev}, \citenamefont {Tu}, \citenamefont {Kretinin},
  \citenamefont {Cao}, \citenamefont {Jalil}, \citenamefont {Withers},
  \citenamefont {Ponomarenko}, \citenamefont {Piot}, \citenamefont {Potemski}
  \emph {et~al.}}]{yu2014hierarchy}%
  \BibitemOpen
  \bibfield  {author} {\bibinfo {author} {\bibfnamefont {G.}~\bibnamefont
  {Yu}}, \bibinfo {author} {\bibfnamefont {R.}~\bibnamefont {Gorbachev}},
  \bibinfo {author} {\bibfnamefont {J.}~\bibnamefont {Tu}}, \bibinfo {author}
  {\bibfnamefont {A.}~\bibnamefont {Kretinin}}, \bibinfo {author}
  {\bibfnamefont {Y.}~\bibnamefont {Cao}}, \bibinfo {author} {\bibfnamefont
  {R.}~\bibnamefont {Jalil}}, \bibinfo {author} {\bibfnamefont
  {F.}~\bibnamefont {Withers}}, \bibinfo {author} {\bibfnamefont
  {L.}~\bibnamefont {Ponomarenko}}, \bibinfo {author} {\bibfnamefont
  {B.}~\bibnamefont {Piot}}, \bibinfo {author} {\bibfnamefont {M.}~\bibnamefont
  {Potemski}},  \emph {et~al.},\ }\href@noop {} {\bibfield  {journal} {\bibinfo
   {journal} {Nature physics}\ }\textbf {\bibinfo {volume} {10}},\ \bibinfo
  {pages} {525} (\bibinfo {year} {2014})}\BibitemShut {NoStop}%
\bibitem [{\citenamefont {Wang}\ \emph {et~al.}(2015)\citenamefont {Wang},
  \citenamefont {Gao}, \citenamefont {Wen}, \citenamefont {Han}, \citenamefont
  {Taniguchi}, \citenamefont {Watanabe}, \citenamefont {Koshino}, \citenamefont
  {Hone},\ and\ \citenamefont {Dean}}]{wang2015evidence}%
  \BibitemOpen
  \bibfield  {author} {\bibinfo {author} {\bibfnamefont {L.}~\bibnamefont
  {Wang}}, \bibinfo {author} {\bibfnamefont {Y.}~\bibnamefont {Gao}}, \bibinfo
  {author} {\bibfnamefont {B.}~\bibnamefont {Wen}}, \bibinfo {author}
  {\bibfnamefont {Z.}~\bibnamefont {Han}}, \bibinfo {author} {\bibfnamefont
  {T.}~\bibnamefont {Taniguchi}}, \bibinfo {author} {\bibfnamefont
  {K.}~\bibnamefont {Watanabe}}, \bibinfo {author} {\bibfnamefont
  {M.}~\bibnamefont {Koshino}}, \bibinfo {author} {\bibfnamefont
  {J.}~\bibnamefont {Hone}}, \ and\ \bibinfo {author} {\bibfnamefont {C.~R.}\
  \bibnamefont {Dean}},\ }\href@noop {} {\bibfield  {journal} {\bibinfo
  {journal} {Science}\ }\textbf {\bibinfo {volume} {350}},\ \bibinfo {pages}
  {1231} (\bibinfo {year} {2015})}\BibitemShut {NoStop}%
\bibitem [{\citenamefont {Kumar}\ \emph {et~al.}(2017)\citenamefont {Kumar},
  \citenamefont {Chen}, \citenamefont {Auton}, \citenamefont {Mishchenko},
  \citenamefont {Bandurin}, \citenamefont {Morozov}, \citenamefont {Cao},
  \citenamefont {Khestanova}, \citenamefont {Shalom}, \citenamefont {Kretinin}
  \emph {et~al.}}]{kumar2017high}%
  \BibitemOpen
  \bibfield  {author} {\bibinfo {author} {\bibfnamefont {R.~K.}\ \bibnamefont
  {Kumar}}, \bibinfo {author} {\bibfnamefont {X.}~\bibnamefont {Chen}},
  \bibinfo {author} {\bibfnamefont {G.}~\bibnamefont {Auton}}, \bibinfo
  {author} {\bibfnamefont {A.}~\bibnamefont {Mishchenko}}, \bibinfo {author}
  {\bibfnamefont {D.~A.}\ \bibnamefont {Bandurin}}, \bibinfo {author}
  {\bibfnamefont {S.~V.}\ \bibnamefont {Morozov}}, \bibinfo {author}
  {\bibfnamefont {Y.}~\bibnamefont {Cao}}, \bibinfo {author} {\bibfnamefont
  {E.}~\bibnamefont {Khestanova}}, \bibinfo {author} {\bibfnamefont {M.~B.}\
  \bibnamefont {Shalom}}, \bibinfo {author} {\bibfnamefont {A.}~\bibnamefont
  {Kretinin}},  \emph {et~al.},\ }\href@noop {} {\bibfield  {journal} {\bibinfo
   {journal} {Science}\ }\textbf {\bibinfo {volume} {357}},\ \bibinfo {pages}
  {181} (\bibinfo {year} {2017})}\BibitemShut {NoStop}%
\bibitem [{\citenamefont {Gorbachev}\ \emph {et~al.}(2014)\citenamefont
  {Gorbachev}, \citenamefont {Song}, \citenamefont {Yu}, \citenamefont
  {Kretinin}, \citenamefont {Withers}, \citenamefont {Cao}, \citenamefont
  {Mishchenko}, \citenamefont {Grigorieva}, \citenamefont {Novoselov},
  \citenamefont {Levitov} \emph {et~al.}}]{gorbachev2014detecting}%
  \BibitemOpen
  \bibfield  {author} {\bibinfo {author} {\bibfnamefont {R.}~\bibnamefont
  {Gorbachev}}, \bibinfo {author} {\bibfnamefont {J.}~\bibnamefont {Song}},
  \bibinfo {author} {\bibfnamefont {G.}~\bibnamefont {Yu}}, \bibinfo {author}
  {\bibfnamefont {A.}~\bibnamefont {Kretinin}}, \bibinfo {author}
  {\bibfnamefont {F.}~\bibnamefont {Withers}}, \bibinfo {author} {\bibfnamefont
  {Y.}~\bibnamefont {Cao}}, \bibinfo {author} {\bibfnamefont {A.}~\bibnamefont
  {Mishchenko}}, \bibinfo {author} {\bibfnamefont {I.}~\bibnamefont
  {Grigorieva}}, \bibinfo {author} {\bibfnamefont {K.}~\bibnamefont
  {Novoselov}}, \bibinfo {author} {\bibfnamefont {L.}~\bibnamefont {Levitov}},
  \emph {et~al.},\ }\href@noop {} {\bibfield  {journal} {\bibinfo  {journal}
  {Science}\ }\textbf {\bibinfo {volume} {346}},\ \bibinfo {pages} {448}
  (\bibinfo {year} {2014})}\BibitemShut {NoStop}%
\bibitem [{\citenamefont {Rivera}\ \emph {et~al.}(2015)\citenamefont {Rivera},
  \citenamefont {Schaibley}, \citenamefont {Jones}, \citenamefont {Ross},
  \citenamefont {Wu}, \citenamefont {Aivazian}, \citenamefont {Klement},
  \citenamefont {Seyler}, \citenamefont {Clark}, \citenamefont {Ghimire} \emph
  {et~al.}}]{rivera2015observation}%
  \BibitemOpen
  \bibfield  {author} {\bibinfo {author} {\bibfnamefont {P.}~\bibnamefont
  {Rivera}}, \bibinfo {author} {\bibfnamefont {J.~R.}\ \bibnamefont
  {Schaibley}}, \bibinfo {author} {\bibfnamefont {A.~M.}\ \bibnamefont
  {Jones}}, \bibinfo {author} {\bibfnamefont {J.~S.}\ \bibnamefont {Ross}},
  \bibinfo {author} {\bibfnamefont {S.}~\bibnamefont {Wu}}, \bibinfo {author}
  {\bibfnamefont {G.}~\bibnamefont {Aivazian}}, \bibinfo {author}
  {\bibfnamefont {P.}~\bibnamefont {Klement}}, \bibinfo {author} {\bibfnamefont
  {K.}~\bibnamefont {Seyler}}, \bibinfo {author} {\bibfnamefont
  {G.}~\bibnamefont {Clark}}, \bibinfo {author} {\bibfnamefont {N.~J.}\
  \bibnamefont {Ghimire}},  \emph {et~al.},\ }\href@noop {} {\bibfield
  {journal} {\bibinfo  {journal} {Nature communications}\ }\textbf {\bibinfo
  {volume} {6}},\ \bibinfo {pages} {6242} (\bibinfo {year} {2015})}\BibitemShut
  {NoStop}%
\bibitem [{\citenamefont {Rivera}\ \emph {et~al.}(2016)\citenamefont {Rivera},
  \citenamefont {Seyler}, \citenamefont {Yu}, \citenamefont {Schaibley},
  \citenamefont {Yan}, \citenamefont {Mandrus}, \citenamefont {Yao},\ and\
  \citenamefont {Xu}}]{rivera2016valley}%
  \BibitemOpen
  \bibfield  {author} {\bibinfo {author} {\bibfnamefont {P.}~\bibnamefont
  {Rivera}}, \bibinfo {author} {\bibfnamefont {K.~L.}\ \bibnamefont {Seyler}},
  \bibinfo {author} {\bibfnamefont {H.}~\bibnamefont {Yu}}, \bibinfo {author}
  {\bibfnamefont {J.~R.}\ \bibnamefont {Schaibley}}, \bibinfo {author}
  {\bibfnamefont {J.}~\bibnamefont {Yan}}, \bibinfo {author} {\bibfnamefont
  {D.~G.}\ \bibnamefont {Mandrus}}, \bibinfo {author} {\bibfnamefont
  {W.}~\bibnamefont {Yao}}, \ and\ \bibinfo {author} {\bibfnamefont
  {X.}~\bibnamefont {Xu}},\ }\href@noop {} {\bibfield  {journal} {\bibinfo
  {journal} {Science}\ }\textbf {\bibinfo {volume} {351}},\ \bibinfo {pages}
  {688} (\bibinfo {year} {2016})}\BibitemShut {NoStop}%
\bibitem [{\citenamefont {Wu}\ \emph {et~al.}(2018)\citenamefont {Wu},
  \citenamefont {Lovorn},\ and\ \citenamefont {MacDonald}}]{wu2018theory}%
  \BibitemOpen
  \bibfield  {author} {\bibinfo {author} {\bibfnamefont {F.}~\bibnamefont
  {Wu}}, \bibinfo {author} {\bibfnamefont {T.}~\bibnamefont {Lovorn}}, \ and\
  \bibinfo {author} {\bibfnamefont {A.}~\bibnamefont {MacDonald}},\ }\href@noop
  {} {\bibfield  {journal} {\bibinfo  {journal} {Physical Review B}\ }\textbf
  {\bibinfo {volume} {97}},\ \bibinfo {pages} {035306} (\bibinfo {year}
  {2018})}\BibitemShut {NoStop}%
\bibitem [{\citenamefont {Yu}\ \emph {et~al.}(2017)\citenamefont {Yu},
  \citenamefont {Liu}, \citenamefont {Tang}, \citenamefont {Xu},\ and\
  \citenamefont {Yao}}]{yu2017moire}%
  \BibitemOpen
  \bibfield  {author} {\bibinfo {author} {\bibfnamefont {H.}~\bibnamefont
  {Yu}}, \bibinfo {author} {\bibfnamefont {G.-B.}\ \bibnamefont {Liu}},
  \bibinfo {author} {\bibfnamefont {J.}~\bibnamefont {Tang}}, \bibinfo {author}
  {\bibfnamefont {X.}~\bibnamefont {Xu}}, \ and\ \bibinfo {author}
  {\bibfnamefont {W.}~\bibnamefont {Yao}},\ }\href@noop {} {\bibfield
  {journal} {\bibinfo  {journal} {Science advances}\ }\textbf {\bibinfo
  {volume} {3}},\ \bibinfo {pages} {e1701696} (\bibinfo {year}
  {2017})}\BibitemShut {NoStop}%
\bibitem [{\citenamefont {Tran}\ \emph {et~al.}(2019)\citenamefont {Tran},
  \citenamefont {Moody}, \citenamefont {Wu}, \citenamefont {Lu}, \citenamefont
  {Choi}, \citenamefont {Kim}, \citenamefont {Rai}, \citenamefont {Sanchez},
  \citenamefont {Quan}, \citenamefont {Singh} \emph
  {et~al.}}]{tran2019evidence}%
  \BibitemOpen
  \bibfield  {author} {\bibinfo {author} {\bibfnamefont {K.}~\bibnamefont
  {Tran}}, \bibinfo {author} {\bibfnamefont {G.}~\bibnamefont {Moody}},
  \bibinfo {author} {\bibfnamefont {F.}~\bibnamefont {Wu}}, \bibinfo {author}
  {\bibfnamefont {X.}~\bibnamefont {Lu}}, \bibinfo {author} {\bibfnamefont
  {J.}~\bibnamefont {Choi}}, \bibinfo {author} {\bibfnamefont {K.}~\bibnamefont
  {Kim}}, \bibinfo {author} {\bibfnamefont {A.}~\bibnamefont {Rai}}, \bibinfo
  {author} {\bibfnamefont {D.~A.}\ \bibnamefont {Sanchez}}, \bibinfo {author}
  {\bibfnamefont {J.}~\bibnamefont {Quan}}, \bibinfo {author} {\bibfnamefont
  {A.}~\bibnamefont {Singh}},  \emph {et~al.},\ }\href@noop {} {\bibfield
  {journal} {\bibinfo  {journal} {Nature}\ ,\ \bibinfo {pages} {1}} (\bibinfo
  {year} {2019})}\BibitemShut {NoStop}%
\bibitem [{\citenamefont {Cao}\ \emph {et~al.}(2018{\natexlab{a}})\citenamefont
  {Cao}, \citenamefont {Fatemi}, \citenamefont {Fang}, \citenamefont
  {Watanabe}, \citenamefont {Taniguchi}, \citenamefont {Kaxiras},\ and\
  \citenamefont {Jarillo-Herrero}}]{cao2018unconventional}%
  \BibitemOpen
  \bibfield  {author} {\bibinfo {author} {\bibfnamefont {Y.}~\bibnamefont
  {Cao}}, \bibinfo {author} {\bibfnamefont {V.}~\bibnamefont {Fatemi}},
  \bibinfo {author} {\bibfnamefont {S.}~\bibnamefont {Fang}}, \bibinfo {author}
  {\bibfnamefont {K.}~\bibnamefont {Watanabe}}, \bibinfo {author}
  {\bibfnamefont {T.}~\bibnamefont {Taniguchi}}, \bibinfo {author}
  {\bibfnamefont {E.}~\bibnamefont {Kaxiras}}, \ and\ \bibinfo {author}
  {\bibfnamefont {P.}~\bibnamefont {Jarillo-Herrero}},\ }\href@noop {}
  {\bibfield  {journal} {\bibinfo  {journal} {Nature}\ }\textbf {\bibinfo
  {volume} {556}},\ \bibinfo {pages} {43} (\bibinfo {year}
  {2018}{\natexlab{a}})}\BibitemShut {NoStop}%
\bibitem [{\citenamefont {Cao}\ \emph {et~al.}(2018{\natexlab{b}})\citenamefont
  {Cao}, \citenamefont {Fatemi}, \citenamefont {Demir}, \citenamefont {Fang},
  \citenamefont {Tomarken}, \citenamefont {Luo}, \citenamefont
  {Sanchez-Yamagishi}, \citenamefont {Watanabe}, \citenamefont {Taniguchi},
  \citenamefont {Kaxiras} \emph {et~al.}}]{cao2018correlated}%
  \BibitemOpen
  \bibfield  {author} {\bibinfo {author} {\bibfnamefont {Y.}~\bibnamefont
  {Cao}}, \bibinfo {author} {\bibfnamefont {V.}~\bibnamefont {Fatemi}},
  \bibinfo {author} {\bibfnamefont {A.}~\bibnamefont {Demir}}, \bibinfo
  {author} {\bibfnamefont {S.}~\bibnamefont {Fang}}, \bibinfo {author}
  {\bibfnamefont {S.~L.}\ \bibnamefont {Tomarken}}, \bibinfo {author}
  {\bibfnamefont {J.~Y.}\ \bibnamefont {Luo}}, \bibinfo {author} {\bibfnamefont
  {J.~D.}\ \bibnamefont {Sanchez-Yamagishi}}, \bibinfo {author} {\bibfnamefont
  {K.}~\bibnamefont {Watanabe}}, \bibinfo {author} {\bibfnamefont
  {T.}~\bibnamefont {Taniguchi}}, \bibinfo {author} {\bibfnamefont
  {E.}~\bibnamefont {Kaxiras}},  \emph {et~al.},\ }\href@noop {} {\bibfield
  {journal} {\bibinfo  {journal} {Nature}\ }\textbf {\bibinfo {volume} {556}},\
  \bibinfo {pages} {80} (\bibinfo {year} {2018}{\natexlab{b}})}\BibitemShut
  {NoStop}%
\bibitem [{\citenamefont {Argentero}\ \emph {et~al.}(2017)\citenamefont
  {Argentero}, \citenamefont {Mittelberger}, \citenamefont {Reza
  Ahmadpour~Monazam}, \citenamefont {Cao}, \citenamefont {Pennycook},
  \citenamefont {Mangler}, \citenamefont {Kramberger}, \citenamefont
  {Kotakoski}, \citenamefont {Geim},\ and\ \citenamefont
  {Meyer}}]{argentero2017unraveling}%
  \BibitemOpen
  \bibfield  {author} {\bibinfo {author} {\bibfnamefont {G.}~\bibnamefont
  {Argentero}}, \bibinfo {author} {\bibfnamefont {A.}~\bibnamefont
  {Mittelberger}}, \bibinfo {author} {\bibfnamefont {M.}~\bibnamefont {Reza
  Ahmadpour~Monazam}}, \bibinfo {author} {\bibfnamefont {Y.}~\bibnamefont
  {Cao}}, \bibinfo {author} {\bibfnamefont {T.~J.}\ \bibnamefont {Pennycook}},
  \bibinfo {author} {\bibfnamefont {C.}~\bibnamefont {Mangler}}, \bibinfo
  {author} {\bibfnamefont {C.}~\bibnamefont {Kramberger}}, \bibinfo {author}
  {\bibfnamefont {J.}~\bibnamefont {Kotakoski}}, \bibinfo {author}
  {\bibfnamefont {A.}~\bibnamefont {Geim}}, \ and\ \bibinfo {author}
  {\bibfnamefont {J.~C.}\ \bibnamefont {Meyer}},\ }\href@noop {} {\bibfield
  {journal} {\bibinfo  {journal} {Nano letters}\ }\textbf {\bibinfo {volume}
  {17}},\ \bibinfo {pages} {1409} (\bibinfo {year} {2017})}\BibitemShut
  {NoStop}%
\bibitem [{\citenamefont {Neek-Amal}\ and\ \citenamefont
  {Peeters}(2014)}]{neek2014graphene}%
  \BibitemOpen
  \bibfield  {author} {\bibinfo {author} {\bibfnamefont {M.}~\bibnamefont
  {Neek-Amal}}\ and\ \bibinfo {author} {\bibfnamefont {F.}~\bibnamefont
  {Peeters}},\ }\href@noop {} {\bibfield  {journal} {\bibinfo  {journal}
  {Applied Physics Letters}\ }\textbf {\bibinfo {volume} {104}},\ \bibinfo
  {pages} {173106} (\bibinfo {year} {2014})}\BibitemShut {NoStop}%
\bibitem [{\citenamefont {Nam}\ and\ \citenamefont
  {Koshino}(2017)}]{nam2017lattice}%
  \BibitemOpen
  \bibfield  {author} {\bibinfo {author} {\bibfnamefont {N.~N.}\ \bibnamefont
  {Nam}}\ and\ \bibinfo {author} {\bibfnamefont {M.}~\bibnamefont {Koshino}},\
  }\href@noop {} {\bibfield  {journal} {\bibinfo  {journal} {Physical Review
  B}\ }\textbf {\bibinfo {volume} {96}},\ \bibinfo {pages} {075311} (\bibinfo
  {year} {2017})}\BibitemShut {NoStop}%
\bibitem [{\citenamefont {Stradi}\ \emph {et~al.}(2012)\citenamefont {Stradi},
  \citenamefont {Barja}, \citenamefont {D{\'\i}az}, \citenamefont {Garnica},
  \citenamefont {Borca}, \citenamefont {Hinarejos}, \citenamefont
  {S{\'a}nchez-Portal}, \citenamefont {Alcam{\'\i}}, \citenamefont {Arnau},
  \citenamefont {De~Parga} \emph {et~al.}}]{stradi2012electron}%
  \BibitemOpen
  \bibfield  {author} {\bibinfo {author} {\bibfnamefont {D.}~\bibnamefont
  {Stradi}}, \bibinfo {author} {\bibfnamefont {S.}~\bibnamefont {Barja}},
  \bibinfo {author} {\bibfnamefont {C.}~\bibnamefont {D{\'\i}az}}, \bibinfo
  {author} {\bibfnamefont {M.}~\bibnamefont {Garnica}}, \bibinfo {author}
  {\bibfnamefont {B.}~\bibnamefont {Borca}}, \bibinfo {author} {\bibfnamefont
  {J.}~\bibnamefont {Hinarejos}}, \bibinfo {author} {\bibfnamefont
  {D.}~\bibnamefont {S{\'a}nchez-Portal}}, \bibinfo {author} {\bibfnamefont
  {M.}~\bibnamefont {Alcam{\'\i}}}, \bibinfo {author} {\bibfnamefont
  {A.}~\bibnamefont {Arnau}}, \bibinfo {author} {\bibfnamefont {A.~V.}\
  \bibnamefont {De~Parga}},  \emph {et~al.},\ }\href@noop {} {\bibfield
  {journal} {\bibinfo  {journal} {Physical Review B}\ }\textbf {\bibinfo
  {volume} {85}},\ \bibinfo {pages} {121404} (\bibinfo {year}
  {2012})}\BibitemShut {NoStop}%
\bibitem [{\citenamefont {Jung}\ \emph {et~al.}(2015)\citenamefont {Jung},
  \citenamefont {DaSilva}, \citenamefont {MacDonald},\ and\ \citenamefont
  {Adam}}]{jung2015origin}%
  \BibitemOpen
  \bibfield  {author} {\bibinfo {author} {\bibfnamefont {J.}~\bibnamefont
  {Jung}}, \bibinfo {author} {\bibfnamefont {A.~M.}\ \bibnamefont {DaSilva}},
  \bibinfo {author} {\bibfnamefont {A.~H.}\ \bibnamefont {MacDonald}}, \ and\
  \bibinfo {author} {\bibfnamefont {S.}~\bibnamefont {Adam}},\ }\href@noop {}
  {\bibfield  {journal} {\bibinfo  {journal} {Nature communications}\ }\textbf
  {\bibinfo {volume} {6}},\ \bibinfo {pages} {6308} (\bibinfo {year}
  {2015})}\BibitemShut {NoStop}%
\bibitem [{\citenamefont {Yankowitz}\ \emph {et~al.}(2018)\citenamefont
  {Yankowitz}, \citenamefont {Jung}, \citenamefont {Laksono}, \citenamefont
  {Leconte}, \citenamefont {Chittari}, \citenamefont {Watanabe}, \citenamefont
  {Taniguchi}, \citenamefont {Adam}, \citenamefont {Graf},\ and\ \citenamefont
  {Dean}}]{yankowitz2018dynamic}%
  \BibitemOpen
  \bibfield  {author} {\bibinfo {author} {\bibfnamefont {M.}~\bibnamefont
  {Yankowitz}}, \bibinfo {author} {\bibfnamefont {J.}~\bibnamefont {Jung}},
  \bibinfo {author} {\bibfnamefont {E.}~\bibnamefont {Laksono}}, \bibinfo
  {author} {\bibfnamefont {N.}~\bibnamefont {Leconte}}, \bibinfo {author}
  {\bibfnamefont {B.~L.}\ \bibnamefont {Chittari}}, \bibinfo {author}
  {\bibfnamefont {K.}~\bibnamefont {Watanabe}}, \bibinfo {author}
  {\bibfnamefont {T.}~\bibnamefont {Taniguchi}}, \bibinfo {author}
  {\bibfnamefont {S.}~\bibnamefont {Adam}}, \bibinfo {author} {\bibfnamefont
  {D.}~\bibnamefont {Graf}}, \ and\ \bibinfo {author} {\bibfnamefont {C.~R.}\
  \bibnamefont {Dean}},\ }\href@noop {} {\bibfield  {journal} {\bibinfo
  {journal} {Nature}\ }\textbf {\bibinfo {volume} {557}},\ \bibinfo {pages}
  {404} (\bibinfo {year} {2018})}\BibitemShut {NoStop}%
\bibitem [{\citenamefont {Choi}\ and\ \citenamefont
  {Choi}(2018)}]{choi2018strong}%
  \BibitemOpen
  \bibfield  {author} {\bibinfo {author} {\bibfnamefont {Y.~W.}\ \bibnamefont
  {Choi}}\ and\ \bibinfo {author} {\bibfnamefont {H.~J.}\ \bibnamefont
  {Choi}},\ }\href@noop {} {\bibfield  {journal} {\bibinfo  {journal} {Physical
  Review B}\ }\textbf {\bibinfo {volume} {98}},\ \bibinfo {pages} {241412}
  (\bibinfo {year} {2018})}\BibitemShut {NoStop}%
\bibitem [{\citenamefont {Cao}\ \emph {et~al.}(2016)\citenamefont {Cao},
  \citenamefont {Luo}, \citenamefont {Fatemi}, \citenamefont {Fang},
  \citenamefont {Sanchez-Yamagishi}, \citenamefont {Watanabe}, \citenamefont
  {Taniguchi}, \citenamefont {Kaxiras},\ and\ \citenamefont
  {Jarillo-Herrero}}]{cao2016superlattice}%
  \BibitemOpen
  \bibfield  {author} {\bibinfo {author} {\bibfnamefont {Y.}~\bibnamefont
  {Cao}}, \bibinfo {author} {\bibfnamefont {J.}~\bibnamefont {Luo}}, \bibinfo
  {author} {\bibfnamefont {V.}~\bibnamefont {Fatemi}}, \bibinfo {author}
  {\bibfnamefont {S.}~\bibnamefont {Fang}}, \bibinfo {author} {\bibfnamefont
  {J.}~\bibnamefont {Sanchez-Yamagishi}}, \bibinfo {author} {\bibfnamefont
  {K.}~\bibnamefont {Watanabe}}, \bibinfo {author} {\bibfnamefont
  {T.}~\bibnamefont {Taniguchi}}, \bibinfo {author} {\bibfnamefont
  {E.}~\bibnamefont {Kaxiras}}, \ and\ \bibinfo {author} {\bibfnamefont
  {P.}~\bibnamefont {Jarillo-Herrero}},\ }\href@noop {} {\bibfield  {journal}
  {\bibinfo  {journal} {Physical review letters}\ }\textbf {\bibinfo {volume}
  {117}},\ \bibinfo {pages} {116804} (\bibinfo {year} {2016})}\BibitemShut
  {NoStop}%
\bibitem [{\citenamefont {San-Jose}\ and\ \citenamefont
  {Prada}(2013)}]{san2013helical}%
  \BibitemOpen
  \bibfield  {author} {\bibinfo {author} {\bibfnamefont {P.}~\bibnamefont
  {San-Jose}}\ and\ \bibinfo {author} {\bibfnamefont {E.}~\bibnamefont
  {Prada}},\ }\href@noop {} {\bibfield  {journal} {\bibinfo  {journal}
  {Physical Review B}\ }\textbf {\bibinfo {volume} {88}},\ \bibinfo {pages}
  {121408} (\bibinfo {year} {2013})}\BibitemShut {NoStop}%
\bibitem [{\citenamefont {Tong}\ \emph {et~al.}(2017)\citenamefont {Tong},
  \citenamefont {Yu}, \citenamefont {Zhu}, \citenamefont {Wang}, \citenamefont
  {Xu},\ and\ \citenamefont {Yao}}]{tong2017topological}%
  \BibitemOpen
  \bibfield  {author} {\bibinfo {author} {\bibfnamefont {Q.}~\bibnamefont
  {Tong}}, \bibinfo {author} {\bibfnamefont {H.}~\bibnamefont {Yu}}, \bibinfo
  {author} {\bibfnamefont {Q.}~\bibnamefont {Zhu}}, \bibinfo {author}
  {\bibfnamefont {Y.}~\bibnamefont {Wang}}, \bibinfo {author} {\bibfnamefont
  {X.}~\bibnamefont {Xu}}, \ and\ \bibinfo {author} {\bibfnamefont
  {W.}~\bibnamefont {Yao}},\ }\href@noop {} {\bibfield  {journal} {\bibinfo
  {journal} {Nature Physics}\ }\textbf {\bibinfo {volume} {13}},\ \bibinfo
  {pages} {356} (\bibinfo {year} {2017})}\BibitemShut {NoStop}%
\bibitem [{\citenamefont {Huang}\ \emph {et~al.}(2018)\citenamefont {Huang},
  \citenamefont {Kim}, \citenamefont {Efimkin}, \citenamefont {Lovorn},
  \citenamefont {Taniguchi}, \citenamefont {Watanabe}, \citenamefont
  {MacDonald}, \citenamefont {Tutuc},\ and\ \citenamefont
  {LeRoy}}]{huang2018topologically}%
  \BibitemOpen
  \bibfield  {author} {\bibinfo {author} {\bibfnamefont {S.}~\bibnamefont
  {Huang}}, \bibinfo {author} {\bibfnamefont {K.}~\bibnamefont {Kim}}, \bibinfo
  {author} {\bibfnamefont {D.~K.}\ \bibnamefont {Efimkin}}, \bibinfo {author}
  {\bibfnamefont {T.}~\bibnamefont {Lovorn}}, \bibinfo {author} {\bibfnamefont
  {T.}~\bibnamefont {Taniguchi}}, \bibinfo {author} {\bibfnamefont
  {K.}~\bibnamefont {Watanabe}}, \bibinfo {author} {\bibfnamefont {A.~H.}\
  \bibnamefont {MacDonald}}, \bibinfo {author} {\bibfnamefont {E.}~\bibnamefont
  {Tutuc}}, \ and\ \bibinfo {author} {\bibfnamefont {B.~J.}\ \bibnamefont
  {LeRoy}},\ }\href@noop {} {\bibfield  {journal} {\bibinfo  {journal}
  {Physical review letters}\ }\textbf {\bibinfo {volume} {121}},\ \bibinfo
  {pages} {037702} (\bibinfo {year} {2018})}\BibitemShut {NoStop}%
\bibitem [{\citenamefont {Pelc}\ \emph {et~al.}(2015)\citenamefont {Pelc},
  \citenamefont {Jask{\'o}lski}, \citenamefont {Ayuela},\ and\ \citenamefont
  {Chico}}]{pelc2015topologically}%
  \BibitemOpen
  \bibfield  {author} {\bibinfo {author} {\bibfnamefont {M.}~\bibnamefont
  {Pelc}}, \bibinfo {author} {\bibfnamefont {W.}~\bibnamefont {Jask{\'o}lski}},
  \bibinfo {author} {\bibfnamefont {A.}~\bibnamefont {Ayuela}}, \ and\ \bibinfo
  {author} {\bibfnamefont {L.}~\bibnamefont {Chico}},\ }\href@noop {}
  {\bibfield  {journal} {\bibinfo  {journal} {Physical Review B}\ }\textbf
  {\bibinfo {volume} {92}},\ \bibinfo {pages} {085433} (\bibinfo {year}
  {2015})}\BibitemShut {NoStop}%
\bibitem [{\citenamefont {Lui}\ \emph {et~al.}(2009)\citenamefont {Lui},
  \citenamefont {Liu}, \citenamefont {Mak}, \citenamefont {Flynn},\ and\
  \citenamefont {Heinz}}]{lui2009ultraflat}%
  \BibitemOpen
  \bibfield  {author} {\bibinfo {author} {\bibfnamefont {C.~H.}\ \bibnamefont
  {Lui}}, \bibinfo {author} {\bibfnamefont {L.}~\bibnamefont {Liu}}, \bibinfo
  {author} {\bibfnamefont {K.~F.}\ \bibnamefont {Mak}}, \bibinfo {author}
  {\bibfnamefont {G.~W.}\ \bibnamefont {Flynn}}, \ and\ \bibinfo {author}
  {\bibfnamefont {T.~F.}\ \bibnamefont {Heinz}},\ }\href@noop {} {\bibfield
  {journal} {\bibinfo  {journal} {Nature}\ }\textbf {\bibinfo {volume} {462}},\
  \bibinfo {pages} {339} (\bibinfo {year} {2009})}\BibitemShut {NoStop}%
\bibitem [{\citenamefont {Xue}\ \emph {et~al.}(2011)\citenamefont {Xue},
  \citenamefont {Sanchez-Yamagishi}, \citenamefont {Bulmash}, \citenamefont
  {Jacquod}, \citenamefont {Deshpande}, \citenamefont {Watanabe}, \citenamefont
  {Taniguchi}, \citenamefont {Jarillo-Herrero},\ and\ \citenamefont
  {LeRoy}}]{xue2011scanning}%
  \BibitemOpen
  \bibfield  {author} {\bibinfo {author} {\bibfnamefont {J.}~\bibnamefont
  {Xue}}, \bibinfo {author} {\bibfnamefont {J.}~\bibnamefont
  {Sanchez-Yamagishi}}, \bibinfo {author} {\bibfnamefont {D.}~\bibnamefont
  {Bulmash}}, \bibinfo {author} {\bibfnamefont {P.}~\bibnamefont {Jacquod}},
  \bibinfo {author} {\bibfnamefont {A.}~\bibnamefont {Deshpande}}, \bibinfo
  {author} {\bibfnamefont {K.}~\bibnamefont {Watanabe}}, \bibinfo {author}
  {\bibfnamefont {T.}~\bibnamefont {Taniguchi}}, \bibinfo {author}
  {\bibfnamefont {P.}~\bibnamefont {Jarillo-Herrero}}, \ and\ \bibinfo {author}
  {\bibfnamefont {B.~J.}\ \bibnamefont {LeRoy}},\ }\href@noop {} {\bibfield
  {journal} {\bibinfo  {journal} {Nature materials}\ }\textbf {\bibinfo
  {volume} {10}},\ \bibinfo {pages} {282} (\bibinfo {year} {2011})}\BibitemShut
  {NoStop}%
\bibitem [{\citenamefont {Neek-Amal}\ \emph {et~al.}(2014)\citenamefont
  {Neek-Amal}, \citenamefont {Xu}, \citenamefont {Qi}, \citenamefont {Thibado},
  \citenamefont {Nyakiti}, \citenamefont {Wheeler}, \citenamefont {Myers-Ward},
  \citenamefont {Eddy~Jr}, \citenamefont {Gaskill},\ and\ \citenamefont
  {Peeters}}]{neek2014membrane}%
  \BibitemOpen
  \bibfield  {author} {\bibinfo {author} {\bibfnamefont {M.}~\bibnamefont
  {Neek-Amal}}, \bibinfo {author} {\bibfnamefont {P.}~\bibnamefont {Xu}},
  \bibinfo {author} {\bibfnamefont {D.}~\bibnamefont {Qi}}, \bibinfo {author}
  {\bibfnamefont {P.}~\bibnamefont {Thibado}}, \bibinfo {author} {\bibfnamefont
  {L.}~\bibnamefont {Nyakiti}}, \bibinfo {author} {\bibfnamefont
  {V.}~\bibnamefont {Wheeler}}, \bibinfo {author} {\bibfnamefont
  {R.}~\bibnamefont {Myers-Ward}}, \bibinfo {author} {\bibfnamefont
  {C.}~\bibnamefont {Eddy~Jr}}, \bibinfo {author} {\bibfnamefont
  {D.}~\bibnamefont {Gaskill}}, \ and\ \bibinfo {author} {\bibfnamefont
  {F.}~\bibnamefont {Peeters}},\ }\href@noop {} {\bibfield  {journal} {\bibinfo
   {journal} {Physical Review B}\ }\textbf {\bibinfo {volume} {90}},\ \bibinfo
  {pages} {064101} (\bibinfo {year} {2014})}\BibitemShut {NoStop}%
\bibitem [{\citenamefont {Gao}\ and\ \citenamefont
  {Huang}(2011)}]{gao2011effect}%
  \BibitemOpen
  \bibfield  {author} {\bibinfo {author} {\bibfnamefont {W.}~\bibnamefont
  {Gao}}\ and\ \bibinfo {author} {\bibfnamefont {R.}~\bibnamefont {Huang}},\
  }\href@noop {} {\bibfield  {journal} {\bibinfo  {journal} {Journal of Physics
  D: Applied Physics}\ }\textbf {\bibinfo {volume} {44}},\ \bibinfo {pages}
  {452001} (\bibinfo {year} {2011})}\BibitemShut {NoStop}%
\bibitem [{\citenamefont {Aitken}\ and\ \citenamefont
  {Huang}(2010)}]{aitken2010effects}%
  \BibitemOpen
  \bibfield  {author} {\bibinfo {author} {\bibfnamefont {Z.~H.}\ \bibnamefont
  {Aitken}}\ and\ \bibinfo {author} {\bibfnamefont {R.}~\bibnamefont {Huang}},\
  }\href@noop {} {\bibfield  {journal} {\bibinfo  {journal} {Journal of Applied
  Physics}\ }\textbf {\bibinfo {volume} {107}},\ \bibinfo {pages} {123531}
  (\bibinfo {year} {2010})}\BibitemShut {NoStop}%
\bibitem [{\citenamefont {Thuermer}\ \emph {et~al.}(2015)\citenamefont
  {Thuermer}, \citenamefont {Foster}, \citenamefont {Bartelt}, \citenamefont
  {Rogge}, \citenamefont {McCarty}, \citenamefont {Dubon},\ and\ \citenamefont
  {Bartelt}}]{thuermer2015real}%
  \BibitemOpen
  \bibfield  {author} {\bibinfo {author} {\bibfnamefont {K.}~\bibnamefont
  {Thuermer}}, \bibinfo {author} {\bibfnamefont {M.~E.}\ \bibnamefont
  {Foster}}, \bibinfo {author} {\bibfnamefont {N.~C.}\ \bibnamefont {Bartelt}},
  \bibinfo {author} {\bibfnamefont {P.~C.}\ \bibnamefont {Rogge}}, \bibinfo
  {author} {\bibfnamefont {K.~F.}\ \bibnamefont {McCarty}}, \bibinfo {author}
  {\bibfnamefont {O.~D.}\ \bibnamefont {Dubon}}, \ and\ \bibinfo {author}
  {\bibfnamefont {N.~C.}\ \bibnamefont {Bartelt}},\ }\href@noop {} {\bibfield
  {journal} {\bibinfo  {journal} {Nature Communications}\ }\textbf {\bibinfo
  {volume} {6}} (\bibinfo {year} {2015})}\BibitemShut {NoStop}%
\bibitem [{\citenamefont {S{\"u}le}\ \emph {et~al.}(2015)\citenamefont
  {S{\"u}le}, \citenamefont {Szendr{\"o}}, \citenamefont {Magda}, \citenamefont
  {Hwang},\ and\ \citenamefont {Tapaszt{\'o}}}]{sule2015nanomesh}%
  \BibitemOpen
  \bibfield  {author} {\bibinfo {author} {\bibfnamefont {P.}~\bibnamefont
  {S{\"u}le}}, \bibinfo {author} {\bibfnamefont {M.}~\bibnamefont
  {Szendr{\"o}}}, \bibinfo {author} {\bibfnamefont {G.~Z.}\ \bibnamefont
  {Magda}}, \bibinfo {author} {\bibfnamefont {C.}~\bibnamefont {Hwang}}, \ and\
  \bibinfo {author} {\bibfnamefont {L.}~\bibnamefont {Tapaszt{\'o}}},\
  }\href@noop {} {\bibfield  {journal} {\bibinfo  {journal} {Nano letters}\
  }\textbf {\bibinfo {volume} {15}},\ \bibinfo {pages} {8295} (\bibinfo {year}
  {2015})}\BibitemShut {NoStop}%
\bibitem [{\citenamefont {Plimpton}(1995)}]{plimpton1995fast}%
  \BibitemOpen
  \bibfield  {author} {\bibinfo {author} {\bibfnamefont {S.}~\bibnamefont
  {Plimpton}},\ }\href@noop {} {\bibfield  {journal} {\bibinfo  {journal}
  {Journal of computational physics}\ }\textbf {\bibinfo {volume} {117}},\
  \bibinfo {pages} {1} (\bibinfo {year} {1995})}\BibitemShut {NoStop}%
\bibitem [{\citenamefont {Los}\ and\ \citenamefont
  {Fasolino}(2003)}]{los2003intrinsic}%
  \BibitemOpen
  \bibfield  {author} {\bibinfo {author} {\bibfnamefont {J.}~\bibnamefont
  {Los}}\ and\ \bibinfo {author} {\bibfnamefont {A.}~\bibnamefont {Fasolino}},\
  }\href@noop {} {\bibfield  {journal} {\bibinfo  {journal} {Physical Review
  B}\ }\textbf {\bibinfo {volume} {68}},\ \bibinfo {pages} {024107} (\bibinfo
  {year} {2003})}\BibitemShut {NoStop}%
\bibitem [{\citenamefont {Kolmogorov}\ and\ \citenamefont
  {Crespi}(2005)}]{kolmogorov2005registry}%
  \BibitemOpen
  \bibfield  {author} {\bibinfo {author} {\bibfnamefont {A.~N.}\ \bibnamefont
  {Kolmogorov}}\ and\ \bibinfo {author} {\bibfnamefont {V.~H.}\ \bibnamefont
  {Crespi}},\ }\href@noop {} {\bibfield  {journal} {\bibinfo  {journal}
  {Physical Review B}\ }\textbf {\bibinfo {volume} {71}},\ \bibinfo {pages}
  {235415} (\bibinfo {year} {2005})}\BibitemShut {NoStop}%
\bibitem [{\citenamefont {Merino}\ \emph {et~al.}(2011)\citenamefont {Merino},
  \citenamefont {Svec}, \citenamefont {Pinardi}, \citenamefont {Otero},\ and\
  \citenamefont {Mart{\'i}n-Gago}}]{merino2011strain}%
  \BibitemOpen
  \bibfield  {author} {\bibinfo {author} {\bibfnamefont {P.}~\bibnamefont
  {Merino}}, \bibinfo {author} {\bibfnamefont {M.}~\bibnamefont {Svec}},
  \bibinfo {author} {\bibfnamefont {A.~L.}\ \bibnamefont {Pinardi}}, \bibinfo
  {author} {\bibfnamefont {G.}~\bibnamefont {Otero}}, \ and\ \bibinfo {author}
  {\bibfnamefont {J.~A.}\ \bibnamefont {Mart{\'i}n-Gago}},\ }\href@noop {}
  {\bibfield  {journal} {\bibinfo  {journal} {Acs Nano}\ }\textbf {\bibinfo
  {volume} {5}},\ \bibinfo {pages} {5627} (\bibinfo {year} {2011})}\BibitemShut
  {NoStop}%
\bibitem [{\citenamefont {Meng}\ \emph {et~al.}(2012)\citenamefont {Meng},
  \citenamefont {Wu}, \citenamefont {Zhang}, \citenamefont {Li}, \citenamefont
  {Du}, \citenamefont {Wang},\ and\ \citenamefont {Gao}}]{meng2012multi}%
  \BibitemOpen
  \bibfield  {author} {\bibinfo {author} {\bibfnamefont {L.}~\bibnamefont
  {Meng}}, \bibinfo {author} {\bibfnamefont {R.}~\bibnamefont {Wu}}, \bibinfo
  {author} {\bibfnamefont {L.}~\bibnamefont {Zhang}}, \bibinfo {author}
  {\bibfnamefont {L.}~\bibnamefont {Li}}, \bibinfo {author} {\bibfnamefont
  {S.}~\bibnamefont {Du}}, \bibinfo {author} {\bibfnamefont {Y.}~\bibnamefont
  {Wang}}, \ and\ \bibinfo {author} {\bibfnamefont {H.}~\bibnamefont {Gao}},\
  }\href@noop {} {\bibfield  {journal} {\bibinfo  {journal} {Journal of
  Physics: Condensed Matter}\ }\textbf {\bibinfo {volume} {24}},\ \bibinfo
  {pages} {314214} (\bibinfo {year} {2012})}\BibitemShut {NoStop}%
\end{thebibliography}%


%

\end{document}